\documentclass[aps,prb,twocolumn,showpacs]{revtex4}
\usepackage{amssymb}
\usepackage{graphicx,amsmath}

\newcommand{\bea}{\begin{eqnarray}}
\newcommand{\eea}{\end{eqnarray}}
\newcommand{\beq}{\begin{equation}}
\newcommand{\eeq}{\end{equation}}
\newcommand{\benu}{\begin{enumerate}}
\newcommand{\enu}{\end{enumerate}}

\newcommand{\om}{\omega}

\newcommand{\ep}{\epsilon}

\newcommand{\dl}{\delta}
\newcommand{\lam}{\lambda}

\newcommand{\ham}{\mathcal{H}}

\newcommand{\ptl}{\partial}
\newcommand{\sgn}{{\rm Sgn}}
\newcommand{\bk}{{\bf k}}
\newcommand{\bq}{{\bf q}}
\newcommand{\bp}{{\bf p}}
\newcommand{\br}{{\bf r}}

\newcommand{\bQ}{{\bf Q}}
\newcommand{\bM}{{\bf M}}
\newcommand{\bm}{{\bf m}}

\begin{document}

\title{
Nesting Induced Large Magnetoelasticity in the Iron Arsenide Systems}
\date{\today}
\author{I. Paul}
\affiliation{
Laboratoire Mat\'{e}riaux et Ph\'{e}nom\`{e}nes Quantiques, Universit\'{e} Paris Diderot-Paris 7 \& CNRS,
UMR 7162, 75205 Paris, France
}

\begin{abstract}

A novel feature of the iron arsenides is the magnetoelastic coupling between the long wavelength
in-plane strains of the lattice and the collective spin fluctuations of the electrons near the magnetic
ordering wavevectors. Here, we study its microscopic origin from an electronic model with nested Fermi
pockets and a nominal interaction. We find the couplings diverge with a power-law as the system is tuned
to perfect nesting. Furthermore, the theory reveals how nematicity is boosted by nesting. These
results are relevant for other systems with nesting driven density wave transitions.

\end{abstract}

\maketitle

\section{Introduction}

A possible source of complexity in correlated metals is the coupling between apparently
unrelated degrees of freedom. The richness that can ensue from it is aptly
demonstrated by the iron arsenide (FeAs) systems
that are being studied intensely for their high temperature superconductivity and for their
intricate non-superconducting phases.~\cite{kamihara,review}
At low doping they
undergo a transition from a tetragonal to orthorhombic crystal structure at temperature $T_S$ (where
$C_4$ symmetry is broken)
followed closely by an antiferromagnetic (AF) transition (where time reversal symmetry is
broken) at $T_N \leq T_S$. The presence
of the two seemingly disparate transitions in close proximity suggests the presence of
magnetoelastic coupling (MEC) between their order
parameters.~\cite{mazin-schmalian,fernandes2010,cano2010,ipaul2011,kuo12,fernandes-meingast} The purpose
of this paper is to study the origin of MEC from a microscopic point
of view, and to argue that Fermi surface nesting enhances their magnitudes dramatically.
The theory also shows how nesting enhances nematicity, which is yet another intriguing and intensely-studied
property of FeAs.

An important band structure feature of these materials, which is well established
both theoretically and experimentally, is the nesting between the circular hole pockets centered
around $(0,0)$ and the elliptic electron pockets centered at $(\pi,0)$ and $(0, \pi)$ of the
Brillouin zone defined by the plane of Fe atoms with 1Fe/cell.~\cite{djsingh08,mazin08,arpes}
Its importance
further underlined by the fact that the AF order involves a nesting wavevector, either
$\bQ_1 = (\pi,0)$ or $\bQ_2 = (0, \pi)$, implying a nesting driven density wave transition
from a paramagnetic metal.~\cite{nesting}

Besides the fact that $T_S$ and $T_N$ track each other closely in the
temperature-doping phase diagram, there are few other indirect evidences of MEC in the FeAs systems.
(1) \emph{Ab initio} calculations show that the electron-phonon coupling strength in
the magnetic state increases by 50\% compared to the paramagnetic one.~\cite{boeri} (2) The magnetic
transition temperature $T_N$ is very sensitive to uniaxial pressure.~\cite{dhital,cano-ipaul,hu-setty}
(3) Applying uniaxial pressure
detwins single crystals in the AF phase.~\cite{detwin}
From the theoretical side, the effects of MEC has been studied
phenomenologically.~\cite{cano2010,ipaul2011,cano-ipaul} It has been
shown that the coupling plays a central role in establishing a universal phase diagram of the
FeAs systems.~\cite{cano2010} In particular, the presence of tricritical points at which the AF transition changes
character from
first to second order were predicted, and it has later been confirmed
experimentally.~\cite{kim-rotundu}
More recently, an \emph{ab initio} study of the effects of uniaxial pressure has interpreted their results
using MEC.~\cite{tomic}
However, to the best of our knowledge, until now there has been no investigation of the microscopic
origin of the MEC.

\section{Model}

In FeAs the MEC coupling between the magnetostructural order parameters and the
associated long wavelength fluctuations can be expressed, using symmetry arguments,
by the effective Hamiltonian
\begin{align}
\label{eq:MEC}
\ham_{ME} =& \sum_{\bq, \bp}  \left\{ \lam_O(q, p) \left[ u_O (\bq)
\bM^{\dagger}_{1,\bq + \bp} \cdot \bM_{1,\bp}
-u_O (\bq^{\prime})
\right. \right.
\nonumber \\
\times& \left.  \bM^{\dagger}_{2,\bq^{\prime} + \bp^{\prime}} \cdot \bM_{2,\bp^{\prime}} \right]
+
\lam_A(q, p)
\left[ u_A (\bq)  \bM^{\dagger}_{1,\bq + \bp} \right.
\nonumber \\
\cdot &
\left. \left.  \bM_{1,\bp}
+ u_A (\bq^{\prime}) \bM^{\dagger}_{2,\bq^{\prime} + \bp^{\prime}} \cdot \bM_{2,\bp^{\prime}} \right] \right\}.
\end{align}
Here $\bM_{\alpha,\bq} \equiv \bM(\bQ_{\alpha} + \bq)$ with $\alpha = (1,2)$ denote the
magnetization
around the ordering wavevectors, $u_O (\bq)$ and $u_A (\bq)$ are the Fourier transforms of the
orthorhombic distortion $u_O(\br) \equiv (\partial_x \rho_x - \partial_y \rho_y)/2$ and the
striction $u_A(\br) \equiv (\partial_x \rho_x + \partial_y \rho_y)/2$ respectively, with $\rho_i(\br)$
being the displacements along $i=(x,y)$ of the Fe atoms from their high temperature tetragonal
equilibrium positions at $\br$.~\cite{landau}
The vectors $(\bq^{\prime}, \bp^{\prime})$ are $\pi/2$ rotations of $(\bq, \bp)$ respectively.
Thus, $\lambda_O(q, p)$ and $\lambda_A(q, p)$ are the orthorhombic- and the striction-
MECs. The $\bq= \bp= 0$ term, in particular, denotes the coupling of the static order parameters.

In this paper we calculate $\lambda_a(q,p)$ for $q, p \ll k_F$, $a= (O, A)$,
from a microscopic model of fermions having
nested Fermi pockets with typical Fermi wavevector $k_F$.
The technical details are given in the Appendix~\ref{appen-a}.
Our main result is that,
as the system approaches perfect nesting, $\lambda_a(q, p)$ diverges with a power-law.
This implies that in nested metals the lattice deformations $u_a(\bq)$ are strongly
coupled to certain \emph{collective} electronic degrees of freedom. This coupling is to be
contrasted with the standard electron-phonon case, where nesting induced phonon anomalies
have weaker logarithmic singularity.~\cite{zhao}
We expect this result to be relevant for other metals that exhibit nesting
induced density wave instabilities~\cite{Peierls} such as certain Cr based alloys,~\cite{chromium}
organic conductors,~\cite{organic} transition metal chalcogens such as NbSe$_3$,~\cite{NbSe3}
and rare earth tellurides.~\cite{rare-earth-tellurides} For FeAs systems this
opens the possibility that $u_a(\bq)$ play a vital role in determining physical
properties such as the superconducting gap structure.

\begin{figure}[!!t]
\begin{center}
\includegraphics[width=5cm,trim=0 0 0 0]{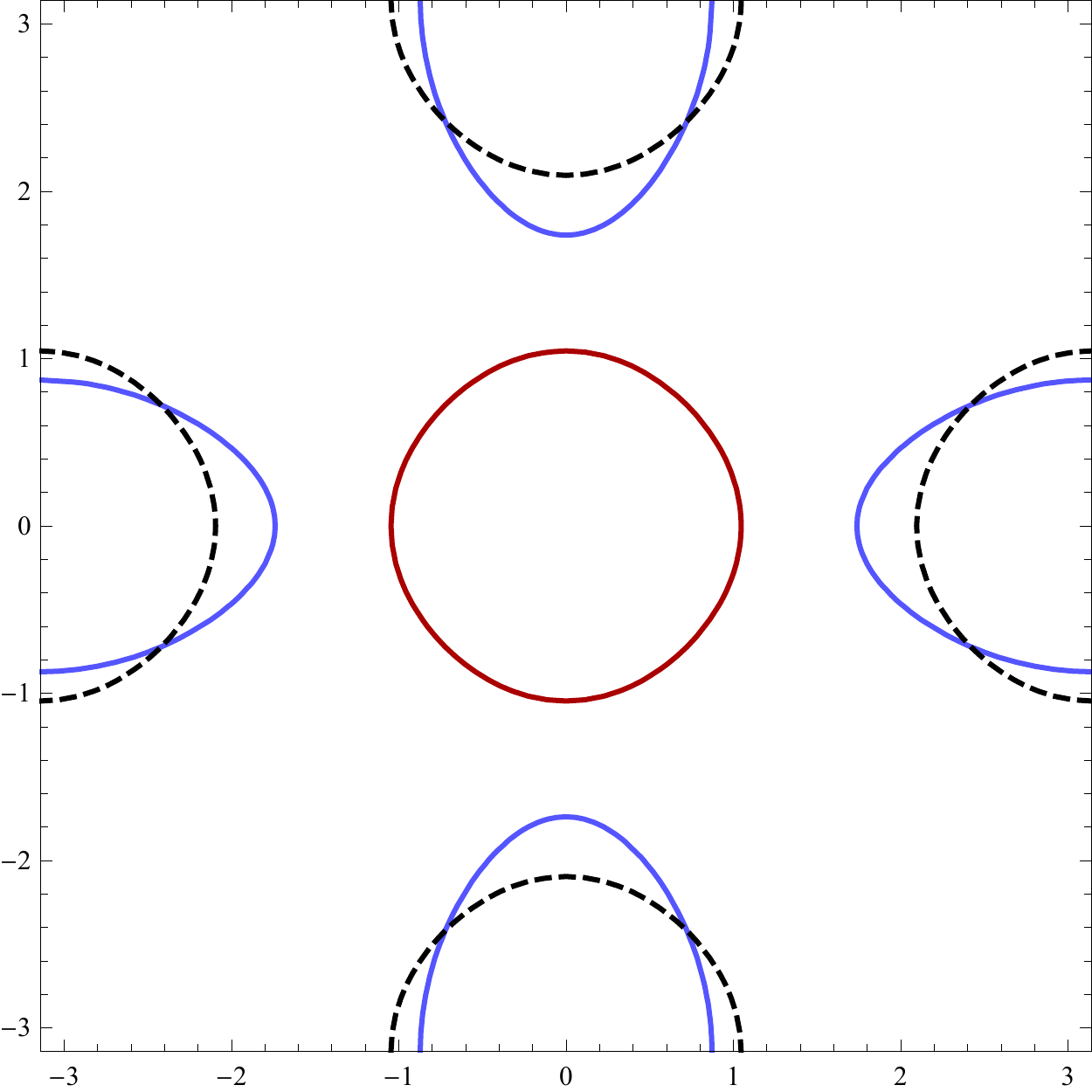}
\caption{
(colour online) Fermi surface topology of the model without lattice distortions
(see Eq.~\ref{eq:h0}).
Nesting between the hole pocket (solid, red) centered at $(0,0)$ and the electron pockets
centered at $(\pi,0)$ and $(0,\pi)$ is tuned by the ellipticity $\eta$ of the latter.
Shown here are $\eta = 0.4$ (solid, blue) and the perfect nesting case with $\eta=0$ (dashed, black).
}
\label{fig1}
\end{center}
\end{figure}

For pedagogical reason we first calculate $\lambda_a(\bq= \bp = 0) \equiv \lambda_a$. For this we consider a
three band model, defined by the Hamiltonian $\ham = \ham_0 + \ham_I$
which has been used in the past for describing the FeAs systems.~\cite{knolle} The band dispersions
are given by
\beq
\label{eq:h0}
\ham_0 = \sum_{\bk, s} \left( \bar{\epsilon}_{\bk}^{\alpha} \alpha^{\dagger}_{\bk, s} \alpha_{\bk, s}
+ \bar{\epsilon}_{\bk}^{\beta} \beta^{\dagger}_{\bk, s} \beta_{\bk, s}
+ \bar{\epsilon}_{\bk}^{\gamma} \gamma^{\dagger}_{\bk, s} \gamma_{\bk, s}
\right),
\eeq
where
\[
\bar{\epsilon}_{\bk}^{n} = \epsilon_0^n + 2(\bar{t}^n_x \cos k_x + \bar{t}^n_y \cos k_y)
\]
with the band index $n = (\alpha, \beta, \gamma)$ and spin index $s$. We take
\begin{align}
\bar{t}^n_x &= t_x^n(1 - p_x^n(u_O + u_A)),
\nonumber \\
\bar{t}^n_y &= t_y^n(1 + p_y^n(u_O - u_A)),
\nonumber
\end{align}
such that $\bar{t}^n_{x/y}$ are the dispersions in the
presence of uniform orthorhombic strain $u_O \equiv u_O(\bq=0)$ and striction $u_A \equiv u_A(\bq=0)$.
We describe the hoppings in the absence of distortions by
$t^{\alpha}_{x/y} = t = 1$~eV, $t_x^{\beta} = t_y^{\gamma} = t (1-\eta)$,
$t_y^{\beta} = t_x^{\gamma} = -t (1+\eta)$, $\ep_0^{\beta} = \ep_0^{\gamma} = -\ep_0^{\alpha} = 3$~eV.
Thus, the $\alpha$-band describes a hole pocket centered at $(0,0)$ and the $\beta$- and $\gamma$- bands
describe electron pockets with ellipticity $\pm \eta$ and centered at $\bQ_1$ and $\bQ_2$ respectively.
In the following we study how the MECs vary with $\eta$, with perfect
nesting at $\eta =0$ (see Fig.~\ref{fig1}).
For describing $\bar{t}^n_{x/y}$ we assume, following Su-Schrieffer-Heeger,~\cite{su1979,heeger1988}
that the changes in the hopping integrals
are proportional to the strain-induced variations of the corresponding bond lengths.
For details of the electron-lattice coupling see Appendix~\ref{appen-a2}
We expect that, in practice,
the proportionality constants $p^n_{x/y}$ depend on the different orbital contents of the
FeAs bands.~\cite{orbital}
Within the current simplified model we take $p_x^{\alpha} = p_x^{\beta}=p_1$,
$p_y^{\beta} = p_x^{\gamma}=p_2$, $p_x^{\beta} = p_y^{\gamma}=p_3$ using $C_4$ symmetry, and we set
$p_1 = p_2 = 2p_3 = 1$. Thus, the less dispersive directions at finite $\eta$ are taken to be
less sensitive to the distortions, thereby simulating the different orbital contents of the nested bands.
Note that, the crucial ingredient here is the nesting $\eta$,
while the other parameters enter the theory as quantitative details.

Next we define
$\hat{\bm}_{1,\bq} = \alpha^{\dagger}_{\bk, s_1} \boldsymbol{\sigma}_{s_1 s_2}
\beta_{\bk + \bQ_1 + \bq, s_2}$
and
$\hat{\bm}_{2,\bq} = \alpha^{\dagger}_{\bk, s_1} \boldsymbol{\sigma}_{s_1 s_2}
\gamma_{\bk + \bQ_2 + \bq, s_2}$, with $\boldsymbol{\sigma}$ denoting Pauli matrices
and sum over repeated indices implied. We introduce the interaction
\beq
\label{eq:ham-I}
\ham_I = -U \sum_{\bq} ( \hat{\bm}^{\dagger}_{1,\bq} \cdot \hat{\bm}_{1,\bq}
+ \hat{\bm}^{\dagger}_{2,\bq} \cdot \hat{\bm}_{2,\bq} ).
\eeq
We take $U=0.07t$ to emphasize the weak coupling nature of the theory.
In fact, the role of the interaction is merely to trigger a
magnetic density wave transition within random phase approximation.

\section{Results}

The derivation of $\lambda_a$, $a=(O,A)$ follows simply from thermodynamic considerations.
The two terms of Eq.~\eqref{eq:ham-I} are decoupled by introducing Hubbard-Stratanovich
fields $(\bM^{\dagger}_{1,\bq}, \bM_{1,\bq})$
and $(\bM^{\dagger}_{2,\bq}, \bM_{2,\bq})$, respectively (see Appendix~\ref{appen-a1}).~\cite{auerbach}
Within random field approximation
the critical magnetic free energy is
$F_M =  U (1- U \chi_{m_1})(\bm_1^2 + \bm_2^2)$ (see Appendix~\ref{appen-a4}).
Here
$\bm_1 = \langle \bM_{1,0} \rangle$ and $\bm_2 = \langle \bM_{2,0} \rangle$ are the magnetic
order parameters, and $\chi_{m_1} \equiv \chi (\bQ_1, \omega =0)$ is the bare static
interband magnetic
susceptibility at $\bQ_1$ obtained from the Fourier transform of
$\chi (\bQ_1+ \bq, \tau) = \langle T_{\tau} \bM^{\dagger}_{1, \bq} (\tau) \cdot \bM_{1, \bq}(0) \rangle/3$,
$T_{\tau}$ being the imaginary time ordering operator.
Next, from the definition of the magnetoelastic free energy
$F_{ME} \equiv \sum_a (\ptl F_M / \ptl u_a) u_a$,~\cite{cano2010} and comparing with Eq.~\ref{eq:MEC}, we get
(see Appendix~\ref{appen-a4})
\beq
\label{eq:lambda1}
\lambda_a = -U^2 \left( \ptl \chi_{m_1} / \ptl u_a \right)_{u_a =0}.
\eeq
Since $\chi_{m_1} \propto \ln (\eta)$ due to the nesting, already from the above Eq.\ we expect that
$\lambda_a \propto 1/\eta$ provided the distortions $u_a$ change the \emph{relative} ellipticity
of the two bands. That this is indeed the case is evident from the expressions for
$\bar{\epsilon}_{\bk}^{n}$.
Here we neglect strain dependence of $U$, since it gives non-singular contribution.

For simplicity we calculate $\lambda_a$ at temperature $T=0$ in the paramagnetic phase,
and later comment about
finite-$T$ effects. In terms of the fermion dispersions we get
\begin{align}
\label{eq:lambda2}
\lambda_a &= 2 U^2 \sum_{\bk} \frac{\ptl \bar{\ep}_{\bk}^{\alpha}}{\ptl u_a}
\left[ \frac{\dl(\ep_{\bk}^{\alpha})}{\ep^{\beta}_{\bk + \bQ_1}}
- \frac{ n_F(\ep_{\bk}^{\alpha})- n_F(\ep^{\beta}_{\bk + \bQ_1})}{(\ep_{\bk}^{\alpha}-\ep^{\beta}_{\bk + \bQ_1})^2}
\right]
\nonumber \\
&+ \alpha \leftrightarrow \beta,
\end{align}
where $\ep_{\bk}^n$ are the undistorted dispersions and $n_F$ is the Fermi function.
In the above the leading contribution is given by the terms with the $\dl$-functions. To calculate
the $\alpha$-band Fermi surface contributions we note that,
on this manifold $\ep^{\beta}_{\bk + \bQ_1}= 2 t \eta (\cos k_x - \cos k_y)$ which has $B_{1g}$ symmetry.
The $(\cos k_x - \cos k_y)$ factor is precisely canceled by $(\ptl \bar{\ep}_{\bk}^{\alpha}/\ptl u_O)$,
the $B_{1g}$ nature of which is guaranteed by the $C_4$ symmetry of the $\alpha$-band.
This gives rise to a singular $1/\eta$ contribution.
Correspondingly, since $(\ptl \bar{\ep}_{\bk}^{\alpha}/\ptl u_A)$ has
$A_{1g}$ symmetry, it does not contribute to the singularity of $\lambda_A$. On the other hand,
as the $\beta$-band is only $C_2$ symmetric, $(\ptl \bar{\ep}_{\bk+ \bQ_1}^{\beta}/\ptl u_a)$
have both $B_{1g}$ and $A_{1g}$ components,
with the former giving singular contributions to both $\lambda_O$ and $\lambda_A$.
We finally get,
\beq
\label{eq:lambda3}
\lambda_{a} = - U^2 \nu_0 l_{a}/(\eta) + \cdots,
\eeq
where the ellipsis (here and henceforth) denote subleading terms,
$\nu_0$ is the density of states of the $\alpha$-band
at the Fermi surface,
and $l_O = (2p_1 - p_2 - p_3)$ and $l_A = (p_2 - p_3)$.
Experimentally, in all the FeAs systems $\lambda_O$ is negative such that
in the AF phase the ferromagnetic bonds are shorter than the antiferromagnetic ones.
In this calculation we get the sign of the singular contribution to be negative
by appropriately choosing $p^n_{x/y}$.
\begin{figure}[!!t]
\begin{center}
\includegraphics[width=8.5cm,trim=0 0 0 0]{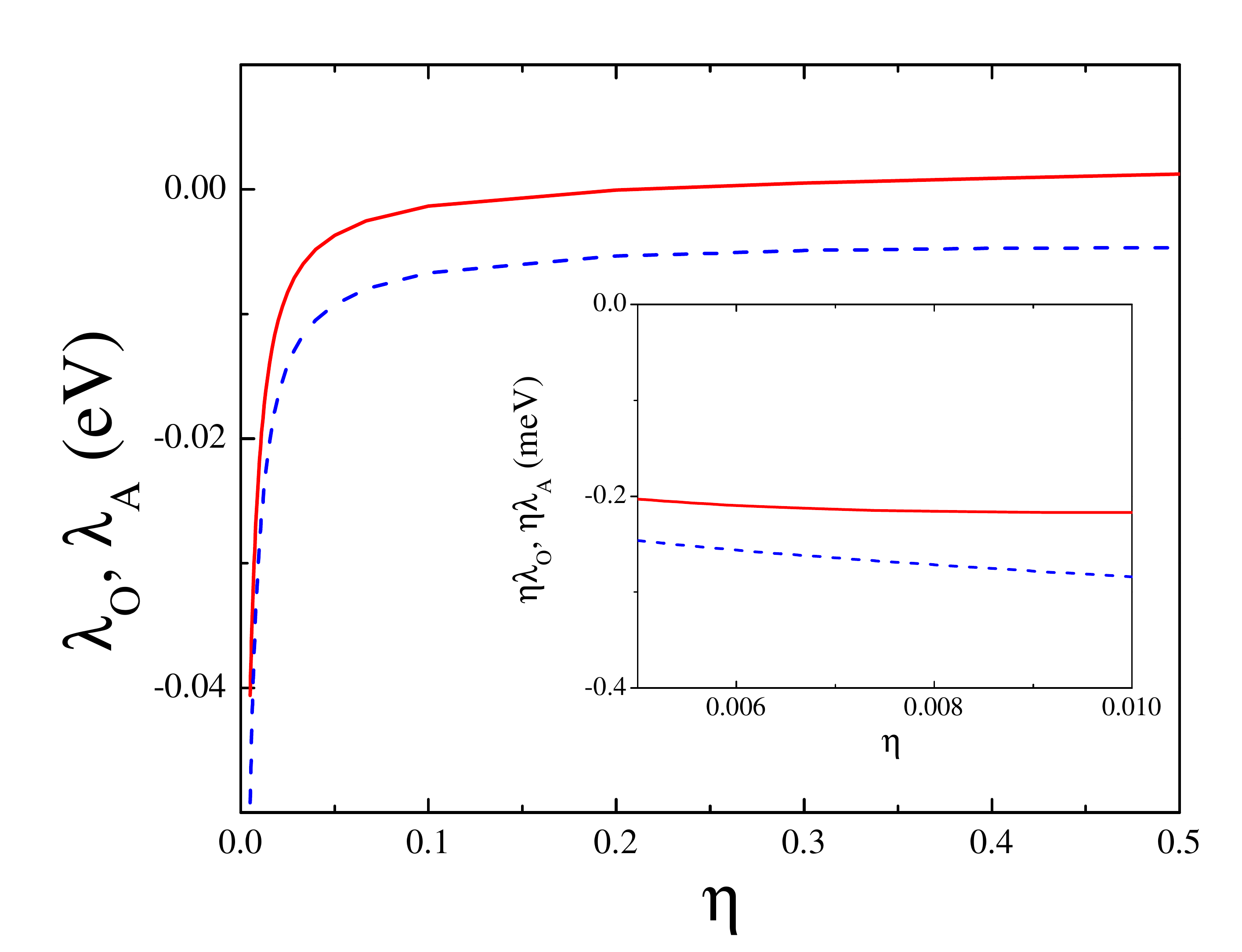}
\caption{
(colour online) Divergence of the orthorhombic and the striction magnetoelastic constants
(defined through Eq.~\ref{eq:MEC}) $\lambda_O$ (solid, red) and $\lambda_A$ (dashed, blue)
as perfect nesting is approached by reducing the ellipticity $\eta$ of the electron pockets.
Inset shows saturation of $\eta \lambda_a$, $a=(O,A)$, at the lowest $\eta$, demonstrating
the $1/\eta$ power-law (see Eq.~\ref{eq:lambda3}).
}
\label{fig2}
\end{center}
\end{figure}

The fact that the remaining terms of Eq.~\ref{eq:lambda2}
are subleading can be understood from the following argument. Near the crossing points
of the two Fermi surfaces (which are potential sources of singularity)
these terms can be expressed as $\sgn(\xi)/\xi^2$,
where $\xi = \ep_{\bk}^{\alpha} - \ep^{\beta}_{\bk + \bQ_1}$. This being odd, the power-law singularity
cancels in the $\xi$-integral. If we take into account $\xi$-dependence of
$(\ptl \bar{\ep}_{\bk}^{\alpha}/\ptl u_a)$ etc, we obtain at most a subleading $\log \eta$ contribution.
Finally, we verified the validity of Eq.~\ref{eq:lambda3} from a direct numerical evaluation of
$\lambda_a$ using Eq.~\ref{eq:lambda2}, as demonstrated in Fig.~\ref{fig2}.

In the above, the importance of the $B_{1g}$ form factor stems from the fact that in the current
model the perfect nesting is achieved by varying the ellipticity $\eta$. If, instead, we set
$\eta=0$ and tune the nesting between ``circular'' bands by varying
$\theta = (|\ep_0^{\alpha}| - |\ep_0^{\beta}|)/|\ep_0^{\alpha}|$, we get singular $1/\theta$ terms that are
associated with $A_{1g}$ form factor. Thus, irrespective of how the nesting is tuned,
the qualitative conclusion, namely the power-law divergence
of $\lambda_a$ remain unchanged. Note, though, that the FeAs bands are closer to
a $\eta$-tuned nesting (at least at low doping), rather than a $\theta$-tuned one where the
two Fermi surfaces (after a $\bQ_1$ shift of the $\beta$-band) do not cross each other.

Next, in order to calculate $\lambda_a(q, p)$, for $q, p \ll k_F$, we generalize the microscopic model of
Eq.~\ref{eq:h0} to
include the coupling between the electrons and the finite-$q$ strains $u_a(\bq)$. This can be done conveniently in real space
through the dependence of the hopping $\bar{t}^n_{x/y}$ to the relative atomic displacements
$(\rho_{x/y}(\br) - \rho_{x/y}(\br^{\prime}))$ between nearest neighbor sites $\br$ and $\br^{\prime}$ (for details see
Appendix~\ref{appen-a2}).
After integrating out the electrons, and
writing explicitly only the singular part of $\lambda_a(q, p)$ we get (see Appendix~\ref{appen-a3})
\begin{align}
\label{eq:lambda5}
\lambda_a(q, p) &=  2 U^2  \sum_{\bk}  \frac{\ptl \bar{\ep}_{\bk}^{\alpha}}{\ptl u_a}
\frac{n_F(\ep_{\bk}^{\alpha}) - n_F(\ep_{\bk - \bq}^{\alpha})}{(\ep_{\bk}^{\alpha} - \ep_{\bk - \bq}^{\alpha})
(\ep_{\bk}^{\alpha} - \ep^{\beta}_{\bk + \bQ_1 + \bp})}
\nonumber \\
 &+ \left( \alpha \leftrightarrow \beta, \bp \rightarrow - \bp, \bq \rightarrow - \bq \right).
\end{align}
In the above the $q$- and $p$- dependencies are quite different.
Since the $q$-dependent factors
$(n_F(\ep_{\bk}^{n}) - n_F(\ep_{\bk + \bq}^{n}))/(\ep_{\bk}^{n} - \ep_{\bk + \bq}^{n})$,
with $n=(\alpha, \beta)$, are strongly peaked at $\ep_{\bk}^{n}=0$, it is
justified to evaluate the remaining parts of the expression on the Fermi surfaces.
For $p=0$ we find that
the $q$-dependence can be expressed by the Lindhard function $\chi_0(q)$ of the $\alpha$-band.
On the other hand, for $q \rightarrow 0$
and $p/k_F < \eta/2$, we find that
the singularity and its pre-factor stays unchanged such that $\lambda_a(0,p) = \lambda_a(0,0)$. In the
opposite limit $p/k_F \gg \eta/2$, the $1/\eta$ singularity is absent.
Taken together, Eq.~\ref{eq:lambda3} can be generalized to
\beq
\label{eq:lambda4}
\lambda_{a}(q, p) = - U^2 \chi_0(q) l_{a}/(\eta) + \cdots,
\eeq
for $p/k_F < \eta/2$ and $q/k_F \ll 1$.
Thus, the coupling between the long wavelength modes involving the acoustic phonons
and the collective spin fluctuations of the
electrons has the same singularity as the coupling between the order parameters, and is
therefore large, even if the bare electron-phonon coupling is weak.

\section{Discussion}

In the following we comment on the implications of the above results for the FeAs systems, and more
generally for nested metals.

(1) Since nesting is the only ingredient, we expect the results to be
relevant for other nested metals.
For e.g., in systems that show nesting induced charge density
wave transitions,~\cite{NbSe3,rare-earth-tellurides}
we expect large coupling between the long wavelength strains and the collective
charge fluctuations.

(2) In the particular context of the FeAs, the relevance of the orthorhombic
MEC $\lambda_O(q)$ has already been pointed out in several phenomenology-based
studies.~\cite{fernandes2010,cano2010,ipaul2011,cano-ipaul}
The current work bolsters these earlier studies
by providing a means to understand why this coupling is large from a microscopic point of view.
We note though, at least in the current simplified
model, the singularities from the two nested bands have opposite signs, and therefore,
for the pre-factor $l_O$ to be non-zero it is crucial that their orbital contents be different,
which is thought to be the
case in FeAs.~\cite{orbital} In practice, the pre-factors $l_a$ in Eq.~\ref{eq:lambda4} need to be evaluated
using \emph{ab initio} tools, which is outside the scope of this work. Experimentally, the
quantity $\lambda_O$ can be obtained from a measurement of the variation of $T_N$ with
orthorhombic strain $u_O$.

(3) The relevance of the striction MEC $\lambda_A(q, p)$ is less obvious for the FeAs, even if
it is large in the current model. Experimentally, $T_N$ is more sensitive to uniaxial rather
than hydrostatic pressure. One reason is that the striction elastic constant
is large in contrast to the orthorhombic one
which is known to be soft in the vicinity of the magneto-structural
transitions. This effectively reduces the effect of hydrostatic pressure.
A second possibility is that the coefficient $l_A$ is small for the FeAs systems. In fact,
instead of taking two electron bands as in the current model, if we consider nesting of the hole band
with a single $C_4$-symmetric electron band,
we find $\lambda_A(q)$ to be non-singular because in this case $p_2 = p_3$.
\begin{figure}[!!t]
\begin{center}
\includegraphics[width=8cm,trim=10 0 0 0]{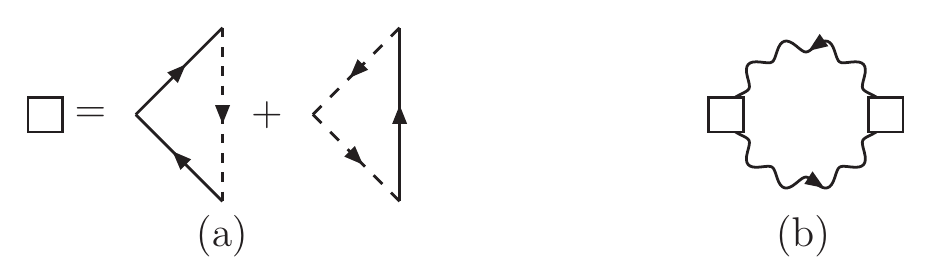}
\caption{
(a) Diagrammatic representation of the magnetoelastic constants
as the sum of fermionic triangles weighted by appropriate form factors (not shown).
Solid and dashed lines imply holes and electrons respectively.
(b) The fermionic triangles enter in the Azlamasov-Larkin graphs for certain correlation
functions (see text), thereby providing nesting induced large contribution. The wavy lines
imply antiferromagnetic spin fluctuations.
}
\label{fig3}
\end{center}
\end{figure}

(4) At finite $T$ we get
$\lambda_a(q, T) \propto 1/{\rm max}[T/\left|\ep_0^{\alpha} \right|, \eta]$, indicating
a $1/T$ dependence at sufficiently large temperature.
The effect of finite lifetime is analogous.

(5) The above results also imply that nesting boosts spin fluctuation induced nematic
softening. This is established most readily from the following argument using diagrams.
The MECs can be represented by fermionic
triangles with weight factors $(\ptl \bar{\ep}_{\bk}^{\alpha}/\ptl u_a)$
and $(\ptl \bar{\ep}_{\bk + \bQ_1}^{\beta}/\ptl u_a)$ respectively (see Fig.~\ref{fig3}(a) and
Figs.~\ref{supp:fig1}, \ref{supp:fig2} of Appendix~\ref{appen-a}).
These triangles, in conjunction with the antiferromagnetic spin fluctuations around the ordering
wavevectors $\bQ_{1/2}$, also enter in the
so-called Azlamasov-Larkin (AL) graphs for nematic
susceptibilities whose vertices have $B_{1g}$ symmetry (see Fig.~\ref{fig3}(b)).
The technical details are given in the Appendix~\ref{appen-b}.
Modeling the spin fluctuations by
$D(\bq, i\nu_n) = 1/(\delta + q^2 + \left| \nu_n \right|)$, where $\delta$ is the mass that
vanishes at magnetic criticality, we get the AL contribution as
\[
\chi_{AL} \sim (1/\eta)^2 \int_{p, \nu_n}^{\prime} D(\bp, i\nu_n)^2,
\]
where the prime denote $\eta \gg q/k_F, \nu/t$. For $\delta \ll \eta^2$ we find
that $\chi_{AL} \sim 1/\eta^2 \log(\eta^2/\delta)$.
Thus, we conclude that in systems where the nesting is $\eta$-tuned,
the AL contribution of the soft spin fluctuations will be enhanced by a factor $1/\eta^2$
for the static response functions of operators that are $B_{1g}$ symmetric under
point group transformations.
This observation provides a rather general means to understand
why spin fluctuations are effective in driving the various
electronic spin-, charge-, and orbital-nematic
softening,~\cite{fernandesSST,charge-nematic,orbital-nematic,fisher}
as well as the softening of the
orthorhombic elastic constant.~\cite{elastic}
Note that, while the importance of the AL contributions has been already emphasized,~\cite{kontani}
the connection with nesting induced singularity has not been made earlier.
Correspondingly, in systems where
$\theta$-tuned nesting is relevant, we expect softening of modes with $A_{1g}$ symmetry.

(6) Finally, phenomenological studies have argued in favor of MEC also in the iron chalcogenide
superconducting systems such as FeTe$_{1-x}$Se$_x$.~\cite{ipaul2011,ipaul-sengupta}
However, just as the magnetic instability
in these systems cannot be understood from a nesting point of view, similarly the current theory
cannot be applied to argue in favor of large MEC in these materials. In other words, the MEC
in the iron chalcogens
is possibly a consequence of strong interaction and stems from the bond length dependence of
the Heisenberg exchanges.
\begin{figure*}
\begin{center}
\includegraphics[width=17cm,trim=0 0 0 0]{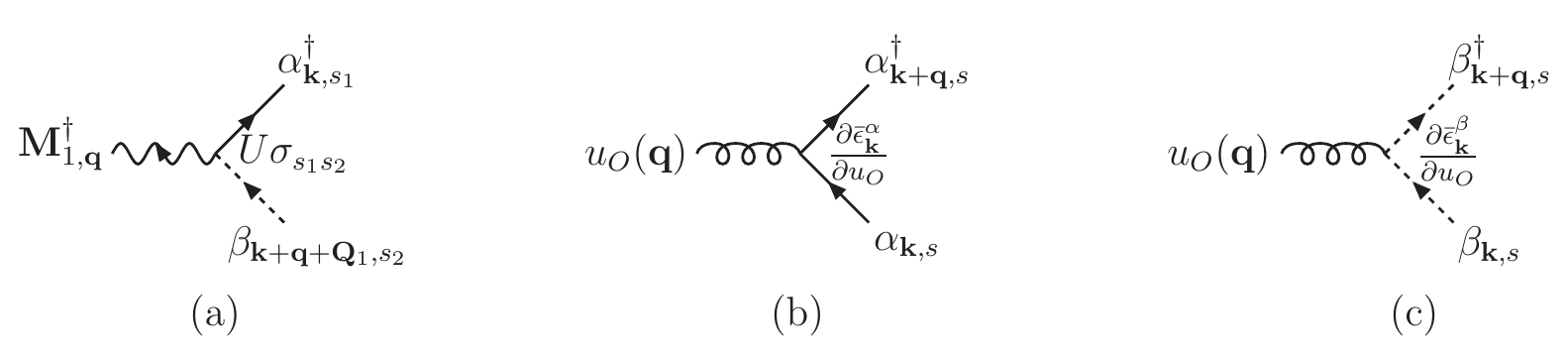}
\caption{
(a) Interaction of spin fluctuations (wavy line) around $(\pi,0)$  with fermions of the $\alpha$-hole (solid line) and
$\beta$-electron (dash line) bands. The interaction involving spin fluctuations around $(0, \pi)$ is similar (not shown).
(b) \& (c) Interaction of orthorhombic strain (gluon line) with fermions. The interaction involving striction is similar
(not shown).
}
\label{supp:fig1}
\end{center}
\end{figure*}

\section{Conclusion}

In the context of the iron arsenide materials, we studied the microscopic origin
of the magnetoelastic couplings between the long wavelength in-plane strains of the lattice
and the collective spin fluctuations of the electrons near the antiferromagnetic ordering
wavevectors. Using a model of electrons with nested Fermi pockets, we find that these
couplings diverge with a power-law as the system approaches perfect nesting. We expect this
singularity to enter the susceptibilities of nematic variables via the Azlamasov-Larkin
contributions. This implies nesting boosts spin fluctuation induced nematic softening.
Moreover, in the future it will be interesting to study if, by means of the magnetoelastic
couplings, the long wavelength strains affect the
superconducting instability within a spin fluctuation mediated pairing scenario.
Finally, our results are relevant for other materials that undergo density wave
instabilities at nesting wavevectors.

\begin{acknowledgements}
The author is very thankful to P. M. R. Brydon, M. Civelli, Y. Gallais, P. Hirschfeld,
C. P\'{e}pin, and C. Timm for insightful discussions.
\end{acknowledgements}

\appendix

\section{}
\label{appen-a}

In this section we provide a detailed derivation of the equations used in the
main text starting from a microscopic model of interacting electrons which are also coupled to
a square lattice. The system is described by the Hamiltonian
\begin{align}
\ham = \ham_{\rm el}^0 + \ham_I + \ham_{\rm el-lattice},
\end{align}
where the first two terms describe the electronic part, and the last term their
coupling to the lattice distortions.

\subsection{Electronic Hamiltonian}
\label{appen-a1}

We use the three-bands model of Ref.~\onlinecite{knolle}, whose dispersions are described by
\begin{align}
\label{supp-ham0}
\ham_{\rm el}^0 = \sum_{\bk, s} \left( \epsilon_{\bk}^{\alpha} \alpha^{\dagger}_{\bk, s} \alpha_{\bk, s}
+ \epsilon_{\bk}^{\beta} \beta^{\dagger}_{\bk, s} \beta_{\bk, s}
+ \epsilon_{\bk}^{\gamma} \gamma^{\dagger}_{\bk, s} \gamma_{\bk, s}
\right),
\end{align}
where $\epsilon_{\bk}^{n} = \epsilon_0^n + 2(t^n_x \cos k_x + t^n_y \cos k_y)$
with the band index $n = (\alpha, \beta, \gamma)$ and spin index $s$.
We describe the hoppings by
$t^{\alpha}_{x/y} = t = 1$~eV, $t_x^{\beta} = t_y^{\gamma} = t (1-\eta)$,
$t_y^{\beta} = t_x^{\gamma} = -t (1+\eta)$, $\ep_0^{\beta} = \ep_0^{\gamma} = -\ep_0^{\alpha} = 3$~eV.
Thus, the $\alpha$-band describes a hole pocket centered at $(0,0)$ and the $\beta$- and $\gamma$- bands
describe electron pockets with ellipticity $\pm \eta$ and centered at $\bQ_1= (\pi,0)$ and
$\bQ_2 = (0, \pi)$ respectively. The associated Fermi surfaces are shown in Fig.~1 of the main text.
The parameter $\eta$ controls the nesting between the hole and the
electron pockets, with $\eta=0$ denoting the idealized situation of perfect nesting.
Note that, in the absence of lattice distortions, $\ham_0$ of Eq.~(2) in the main text coincides with
$\ham_{\rm el}^0$ defined above.
\begin{figure*}
\begin{center}
\includegraphics[width=17cm,trim=0 0 0 0]{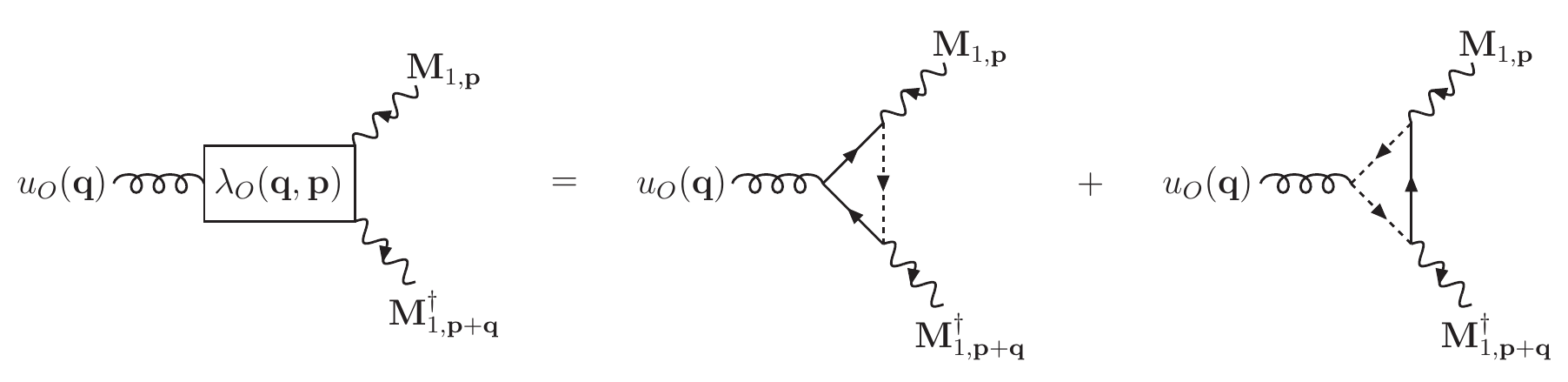}
\caption{
Diagrammatic representation of the magnetoelastic coupling between the spin fluctuations (wavy line) around
$(\pi,0)$ and orthorhombic strain (gluon line). The couplings involving spin fluctuations around $(0, \pi)$
and striction are similar (not shown). The magnetoelastic couplings are three fermion excitations.
}
\label{supp:fig2}
\end{center}
\end{figure*}

Next we define the interband spin operators
\begin{align}
\hat{\bm}_{1,\bq} &= \alpha^{\dagger}_{\bk, s_1} \boldsymbol{\sigma}_{s_1 s_2}
\beta_{\bk + \bQ_1 + \bq, s_2},
\nonumber \\
\hat{\bm}_{2,\bq} &= \alpha^{\dagger}_{\bk, s_1} \boldsymbol{\sigma}_{s_1 s_2}
\gamma_{\bk + \bQ_2 + \bq, s_2}
\nonumber
\end{align}
with sum over repeated indices implied, and we introduce the interaction
\[
\ham_I = -U \sum_{\bq} ( \hat{\bm}^{\dagger}_{1,\bq} \cdot \hat{\bm}_{1,\bq}
+ \hat{\bm}^{\dagger}_{2,\bq} \cdot \hat{\bm}_{2,\bq} ).
\]
This is Eq.~(3) in the main text.
As noted there, the role of this interaction is only to trigger a spin density
wave transition. We decouple the two interaction terms, which are quartic in fermion
variables, by introducing the bosonic Hubbard-Stratanovich fields $(\bM^{\dagger}_{1,\bq}, \bM_{1,\bq})$
and $(\bM^{\dagger}_{2,\bq}, \bM_{2,\bq})$, respectively. After standard steps,~\cite{auerbach}
the interactions can be re-expressed as
\begin{align}
\label{supp:ham-I}
\ham_I &= U \sum_{\bq} \left(
\bM^{\dagger}_{1,\bq} \cdot \bM_{1,\bq} + \bM^{\dagger}_{2,\bq} \cdot \bM_{2,\bq}
\right. \nonumber \\
&+ \bM^{\dagger}_{1,\bq} \cdot \alpha^{\dagger}_{\bk, s_1} \boldsymbol{\sigma}_{s_1 s_2}
\beta_{\bk + \bQ_1 + \bq, s_2}
\nonumber \\
&+ \left.
\bM^{\dagger}_{2,\bq} \cdot \alpha^{\dagger}_{\bk, s_1} \boldsymbol{\sigma}_{s_1 s_2}
\gamma_{\bk + \bQ_2 + \bq, s_2} + {\rm h.c.} \right).
\end{align}
Formally, in the above, $\ham_I$ is quadratic in the fermion variables. The coupling between the
bosonic Hubbard-Stratanovich fields and the fermions is shown graphically in Fig.~\ref{supp:fig1} (a).

\subsection{Electron-Lattice Coupling}
\label{appen-a2}

We express the electron-lattice coupling as
\begin{align}
\ham_{\rm el-lattice} &= - \sum_{\br, s} \left( \rho_x (\br + \hat{x}) - \rho_x (\br) \right)
\left( t^{\alpha}_x p^{\alpha}_x \alpha^{\dagger}_{\br + \hat{x}, s} \alpha_{\br, s}
\right. \nonumber \\
&+ \left.
t^{\beta}_x p^{\beta}_x \beta^{\dagger}_{\br + \hat{x}, s} \beta_{\br, s}
+ t^{\gamma}_x p^{\gamma}_x \gamma^{\dagger}_{\br + \hat{x}, s} \gamma_{\br, s} + {\rm h.c.}
\right)
\nonumber \\
&+ x \rightarrow y.
\end{align}
In the above $\bf{\rho} (\br)$ are the atomic displacements from their tetragonal equilibrium
positions at $\br$. This is a two-dimensional and multiband generalization of the
Su-Schrieffer-Heeger coupling introduced to study conducting polymers.~\cite{su1979,heeger1988}
Physically, it implies that the hoppings parameters depend on the bond-lengths, and that their
variations are proportional to the variations in the bond-lengths. Here, $(p^n_x, p^n_y)$ with
$n = (\alpha, \beta, \gamma)$ are the band- and direction- dependent proportionality constants.
They depend on the orbital contents of the bands, and, in practice, should be obtained using
\emph{ab initio} methods. In the current simplified model we take $p_x^{\alpha} = p_x^{\alpha}=p_1$,
$p_y^{\beta} = p_x^{\gamma}=p_2$, $p_x^{\beta} = p_y^{\gamma}=p_3$ using $C_4$ symmetry, and we set
$p_1 = p_2 = 2p_3 = 1$. Next, we Fourier transform the above Eq.\, and we expand in small $q$ since
we are interested only in the coupling to the acoustic phonons and the uniform strains.
Using the definition of the orthorhombic strain $u_O(\br) \equiv (\partial_x \rho_x - \partial_y \rho_y)/2$
and that of the striction $u_A(\br) \equiv (\partial_x \rho_x + \partial_y \rho_y)/2$, and their
Fourier transforms $u_O(\bq) = i(q_x \rho_x(\bq) - q_y \rho_y(\bq))/2$,
and $u_A(\bq) = i(q_x \rho_x(\bq) + q_y \rho_y(\bq))/2$, we get
\begin{align}
\label{supp:el-lattice}
\ham_{\rm el-lattice} &= - 2 \sum_{\bq, \bk, s} \left[ t_x^{\alpha} p_x^{\alpha} \left(
u_O(\bq) + u_A(\bq) \right) \cos k_x
\right. \nonumber \\
&+ \left.
t_y^{\alpha} p_y^{\alpha} \left(
u_A(\bq) - u_O(\bq) \right) \cos k_y \right] \alpha^{\dagger}_{\bk + \bq, s} \alpha_{\bk, s}
\nonumber \\
&+ (\alpha \rightarrow \beta) + (\alpha \rightarrow \gamma).
\end{align}
In the limit $q \rightarrow 0$, where only the uniform orthorhombic and striction strains
$u_O$ and $u_A$ are present, the hopping parameters are modified compared to Eq.~\eqref{supp-ham0}
such that
\begin{align}
\label{supp:bar-t}
t_x^n &\rightarrow \bar{t}_x^n = t_x^n \left( 1 - p_x^n(u_O + u_A) \right),
\nonumber \\
t_y^n &\rightarrow \bar{t}_y^n = t_y^n \left( 1 + p_y^n(u_O - u_A) \right).
\end{align}
The corresponding dispersions $\bar{\epsilon}_{\bk}^n$ in the presence of uniform strains, where
\begin{align}
\label{supp:bar-epsilon}
\bar{\epsilon}_{\bk}^{n} = \epsilon_0^n + 2(\bar{t}^n_x \cos k_x + \bar{t}^n_y \cos k_y),
\end{align}
are described in Eq.~(2) of the main text. Using Eq.~\eqref{supp:bar-epsilon} we can re-write Eq.~\eqref{supp:el-lattice}
as
\begin{align}
\label{supp:el-lattice2}
\ham_{\rm el-lattice} &= \sum_{\bq, \bk, s} \left[ \frac{\partial \bar{\epsilon}_{\bk}^{\alpha}}{\partial u_O} u_O (\bq)
+ \frac{\partial \bar{\epsilon}_{\bk}^{\alpha}}{\partial u_A} u_A (\bq) \right]
\alpha^{\dagger}_{\bk + \bq, s} \alpha_{\bk, s}
\nonumber \\
&+ (\alpha \rightarrow \beta) + (\alpha \rightarrow \gamma).
\end{align}
Examples of the above coupling is shown graphically in Fig.~\ref{supp:fig1} (b) and (c).

\begin{figure*}
\begin{center}
\includegraphics[width=17cm,trim=0 0 0 0]{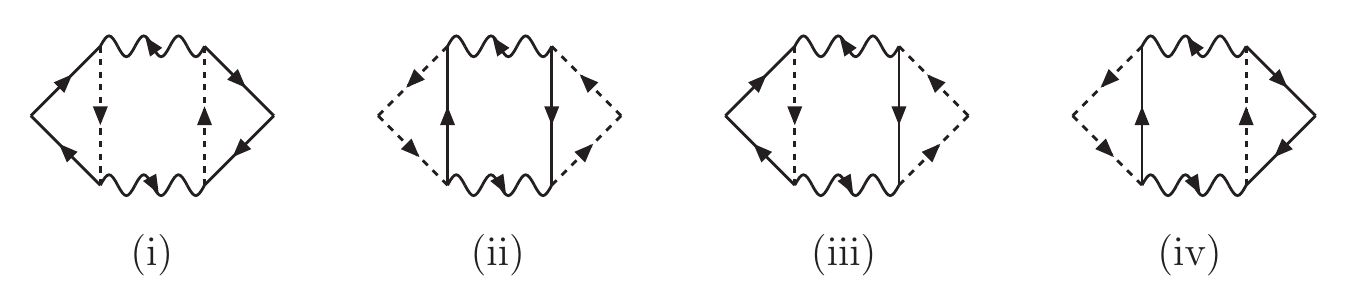}
\caption{
Azlamasov-Larkin graphs representing contribution of spin fluctuations around
$(\pi,0)$ to the charge nematic susceptibility. There is similar contribution
of the spin fluctuations around $(0, \pi)$ (not shown). The triangular vertices
of the graphs, like the magnetoelatsic couplings, are boosted by nesting.
}
\label{supp:fig3}
\end{center}
\end{figure*}

\subsection{Magnetoelastic Coupling}
\label{appen-a3}

The magnetoelastic coupling is obtained by integrating out the fermions. This is possible
because both the electron-electron interaction in Eq.~\eqref{supp:ham-I} and the electron-lattice
interaction in Eq.~\eqref{supp:el-lattice2} are formally quadratic in the fermion variables.
The result of this step is most readily understood in the diagrammatic language. Thus, the
antiferromagnetic spin fluctuations (wavy lines) and the lattice distortions (gluon lines) are
connected by internal fermion loops in Fig.~\ref{supp:fig2}. When the external frequencies are set
to zero, the lowest order terms define an effective magnetoelatsic Hamiltonian
\begin{align}
\ham_{ME} &= \sum_{\bq, \bp}  \left\{ \lambda_O(q, p) u_O (\bq) + \lambda_A(q,p) u_A(\bq) \right\}
\nonumber \\
&\times
\bM^{\dagger}_{1,\bq + \bp} \cdot \bM_{1,\bp} + \cdots,
\end{align}
where the ellipses denote equivalent symmetry-related terms involving $(\bM^{\dagger}_{2,\bp} , \bM_{2,\bp})$.
Note that, this is Eq.~(1) in the main text. It is obvious from Fig.~\ref{supp:fig2} that the
magnetoelastic couplings $\lambda_a(q, p)$, with $a= (O,A)$, are nothing but convolutions of three fermion
excitations. Thus,
\begin{align}
\label{supp:lambda1}
\lambda_a(q, p) &= \frac{2 U^2}{\beta} \sum_{\bk, \om_n} \frac{\partial \bar{\epsilon}_{\bk}^{\alpha}}
{\partial u_a} G_{\alpha} (\bk, i\om_n) G_{\alpha}(\bk - \bq, i \om_n)
\nonumber \\
&\times
G_{\beta}(\bk + \bp+\bQ_1, i \om_n)
\nonumber \\
&+ \{\alpha \leftrightarrow \beta, (\bp, \bq) \rightarrow (- \bp, -\bq) \}.
\end{align}
Here $\beta$ is inverse temperature, and the fermion Green's functions (in the absence of distortion) are given by
$G_{\alpha / \beta}(\bk, i \om_n) = 1/(i \om_n - \epsilon_{\bk}^{\alpha / \beta})$. After the frequency summation,
and keeping only the leading terms we get
\begin{align}
\label{supp:lambda2}
\lambda_a(q, p)  &=  2 U^2  \sum_{\bk}  \frac{\ptl \bar{\ep}_{\bk}^{\alpha}}{\ptl u_a}
\frac{n_F(\ep_{\bk}^{\alpha}) - n_F(\ep_{\bk - \bq}^{\alpha})}{(\ep_{\bk}^{\alpha} - \ep_{\bk - \bq}^{\alpha})
(\ep_{\bk}^{\alpha} - \ep^{\beta}_{\bk + \bQ_1 + \bp})}
\nonumber \\
 &+ \{\alpha \leftrightarrow \beta, (\bp, \bq) \rightarrow (- \bp, -\bq) \},
\end{align}
which in the main text is Eq.~\ref{eq:lambda5}. Setting the external momenta to zero, and at zero
temperature we get
\beq
\label{supp:lambda3}
\lambda_a  \equiv \lambda_a(\bq=\bp=0) = 2 U^2 \sum_{\bk} \frac{\ptl \bar{\ep}_{\bk}^{\alpha}}{\ptl u_a}
\frac{\dl(\ep_{\bk}^{\alpha})}{\ep^{\beta}_{\bk + \bQ_1}} + \{\alpha \leftrightarrow \beta \},
\eeq
which are the leading terms (and the only ones of interest) in Eq.~\eqref{eq:lambda2} of the main text. The evaluation of
the r.h.s. of Eq.~\eqref{supp:lambda2} for small external momenta is described in the main text.
This gives Eq.~\eqref{eq:lambda4} of the main text, and, as a limiting case, Eq.~\eqref{eq:lambda3} of the main text.

\subsection{Derivation of $\lambda_a$ from Thermodynamics}
\label{appen-a4}

Note that, $\lambda_a$ (i.e., with zero external momenta) defines the coupling between the order
parameters of the magnetostructural transitions. Consequently, it is possible to obtain Eq.~\eqref{supp:lambda3}
from thermodynamic arguments, without resorting to diagrams. This is the path followed in the main text.
We repeat them here, for the sake of completeness, and also fill in with few details. After the
Hubbard-Stratanovich transformation the critical part of the magnetic free energy to one loop order
(which is equivalent to random phase approximation) can be written as
$F_M =  U (1 - U \chi_{m_1})(\bm_1^2 + \bm_2^2)$. Here
$\bm_1 = \langle \bM_{1,0} \rangle$ and $\bm_2 = \langle \bM_{2,0} \rangle$ are the magnetic
order parameters, and $\chi_{m_1} \equiv \chi (\bQ_1, \omega =0)$ is the bare static
interband magnetic
susceptibility at $\bQ_1$ obtained from the Fourier transform of
$\chi (\bQ_1+ \bq, \tau) = \langle T_{\tau} \bM^{\dagger}_{1, \bq} (\tau) \cdot \bM_{1, \bq}(0) \rangle/3$,
$T_{\tau}$ being the imaginary time ordering operator. We get
$\chi_{m_1} = 2\sum_{\bk} (n_F(\bar{\epsilon}_{\bk}^{\alpha}) - n_F(\bar{\epsilon}_{\bk + \bQ_1}^{\beta}))/
(\bar{\epsilon}_{\bk + \bQ_1}^{\beta} - \bar{\epsilon}_{\bk}^{\alpha})$. Next, purely from symmetry
argument, the magnetoelastic free energy can be written as
\beq
F_{ME} = \lambda_O u_O (\bm_1^2 - \bm_2^2) + \lambda_A u_A (\bm_1^2 + \bm_2^2).
\eeq
Since $F_{ME}$ is linear in the strains, it can be considered as a linear response of $F_M$
in the presence of strains. This suggests that $F_{ME} \equiv \sum_a (\ptl F_M / \ptl u_a) u_a$, from which
we get Eqs.~(4) and (5) of the main text.

\section{}
\label{appen-b}

In this section we give further detail concerning the fifth point of the ``discussion''
which relates nematicity with magnetoelasticity. In particular, we show that, in complete
analogy with magnetoelasticity, nesting of hole and electron pockets boosts critical
spin fluctuation driven nematicity.

For the sake of concreteness we consider the charge nematic operator, which for the current model is given by
\[
\hat{\mathcal{O}} \equiv \sum_{\bk , s} (\cos k_x - \cos k_y)
(\alpha^{\dagger}_{\bk, s} \alpha_{\bk, s} + \beta^{\dagger}_{\bk, s} \beta_{\bk, s}
+ \gamma^{\dagger}_{\bk, s} \gamma_{\bk, s}).
\]
We define
the static nematic susceptibility as $\chi_n \equiv \int_0^{\beta} d \tau \langle \hat{\mathcal{O}}(\tau)
\hat{\mathcal{O}}(0) \rangle$. In particular, we concentrate on the effect of critical spin fluctuations
on $\chi_n$. The relevant diagrams involving spin fluctuations $(\bM^{\dagger}_{1,\bq} , \bM_{1,\bq})$
and the $\alpha$ and $\beta$ bands are shown in Fig.~\ref{supp:fig3}. Those involving
$(\bM^{\dagger}_{2,\bq} , \bM_{2,\bq})$ and the $\alpha$ and $\gamma$ bands give a factor 2 (not shown).
These are the so-called Azlamasov-Larkin (AL) graphs. Writing the spin fluctuation propagator as
$D(\bq, i\nu_n) = 1/(\delta + q^2 + \left| \nu_n \right|)$, where $\delta$ is the mass that
vanishes at magnetic quantum criticality, we get the AL contribution
\beq
\chi_n^{AL} \propto \frac{1}{\beta} \sum_{\bp, \nu_n} \Lambda(\bp, i \nu_n, i \nu_n)^2 D(\bp, i \nu_n)^2,
\eeq
where
\begin{align}
\label{supp:Lambda}
\Lambda(\bp, i \nu_n, i \nu_n) &= \frac{1}{\beta} \sum_{\bk, \om_n} (\cos k_x - \cos k_y)
G_{\alpha}(\bk, i \om_n)^2
\nonumber \\
&\times
G_{\beta}(\bk + \bp+\bQ_1, i \om_n + i \nu_n)
\nonumber \\
&+ \{\alpha \leftrightarrow \beta, (\nu_n, \bq) \rightarrow (- \nu_n, -\bq) \}.
\end{align}
On comparing Figs.~\ref{supp:fig2} and \ref{supp:fig3}, and Eqs.~\eqref{supp:lambda1} and
\eqref{supp:Lambda}, it is clear that $\Lambda(\bp, i \nu_n, i \nu_n)$, like the magnetoelatsic couplings
$\lambda_a(q,p)$, are convolutions of three-fermion excitations.
Furthermore, noting that the $1/\eta$ singularity of $\lambda_a(q,p)$
is related to $B_{1g}$ form factor, we conclude that for $(p/k_F, \nu_n/t) \ll \eta$,
$\Lambda(\bp, i \nu_n, i \nu_n) \propto 1/\eta$. In two space dimensions and for $\delta \ll \eta^2$ this
eventually leads to the result quoted in the main text, namely $\chi_n^{AL} \sim 1/\eta^2 \log(\eta^2/\delta)$.

\end{document}